\documentclass[%
reprint,
 aps,
pra,
]{revtex4-2}

\usepackage{graphicx}
\usepackage{dcolumn}
\usepackage{float}
\usepackage{bm}
\usepackage{multirow}
\usepackage{caption}
\usepackage{blindtext}
\usepackage[dvipsnames]{xcolor}
\usepackage{graphicx}
\usepackage{subfigure}
\usepackage{appendix}
\begin{document}

\title{{Precise charge state distribution of projectile ions through solid targets }}
\author{Manpreet Kaur$^1$, Sanjeev Kumar$^1$ and T. Nandi$^{2*}$}
\affiliation{$^{1}$Department of Physics, University Institute of Sciences, Chandigarh University, Gharuan, Mohali, Punjab 140413, India}
\affiliation{$^{2}$Department of Physics, Ramakrishna Mission Vivekananda Educational and Research Institute, PO Belur Math, Dist Howrah 711202, West Bengal, India}
\thanks {Email:\hspace{0.0cm} nanditapan@gmail.com (corresponding author). Past address: Inter University Accelerator Centre, JNU New Campus, Aruna Asaf Ali Marg, New Delhi-110067, India.}
\begin{abstract}
The charge state distribution (CSD) of the projectile ions through solid targets in the intermediate energy range (1 MeV/u $<$ E $<$ 4 MeV/u) has a major impact on the collision of the ion atom and accelerator physics. We explore the mean charge states taken from the empirical formula [Schiwietz $et~al.$, Nucl. Inst. Meths. {\bf 225}, 4(2004)] are only good for projectile ions with $Z_1 \le 16$. To solve this issue,  we develop a model in which instead of a single formula, if we employ four formulae, the comparative picture between experimental and empirical data becomes impressive. Furthermore, the CSDs with the mean charge state so obtained and the Gaussian distribution function having distribution width given by [Novikov and Teplove, Phys. Lett. {\bf378}, 1286(2014)] compare well with the experimentally measured CSDs for the entire range of projectile ions. We believe that precise CSDs will be highly useful in both ion-atom collision and accelerator physics.
\end{abstract}
 
\maketitle

\section{Introduction}
When a monochromatic focused ion beam with a particular initial charge state (q) passes through a solid target, it undergoes atomic excitation and ionization processes, which lead to a rapid change in the projectile charge state inside the target ($q^i_m$). Due to these atomic processes, the projectile ion loses some of its kinetic energy inside the target, and the energy loss depends on the kinetic energy of the ion beam, the atomic numbers of the projectile ($Z_1$), the target ($Z_2$) and the initial charge state of the projectile ions. The final interaction of the projectile ions with the target surface layers takes place at the emergent energy of the projectile ions, which experience a profound charge exchange \cite{sharma2019disentangling}. Thus, this charge exchange plays an important role in attaining the final charge states outside the target ($q_m^{o})$ with which the projectile ions emerge from the target \cite{sharma2019disentangling}. Note that $q^i_m$ is higher than $q^o_m$ \cite{chatterjee2021significance, chatterjee2022understanding, PhysRevA.108.052817} because the projectile ions pass through the beam-foil plasma \cite{sharma2016experimental} formed inside the foil by ion-solid interactions. In contrast, the projectile ions with $q^i_m$ have to pass through the exit layers of the target foil from where the quasifree electrons present at the surface of the foil can easily jump to the projectile ions. The $q^o_m$ and the charge-state distribution (CSD) of the emerging projectile ions are accurately measured by standard electromagnetic methods \cite{MAIDIKOV1982295}.

\par
The prediction of $q^o_m$ is also performed by empirical models if $q^o_m$ is reached to charge equilibrium. When we consider charge equilibrium cases, the initial charge state of the projectile ions does not play any role in estimating the $q^o_m$ and CSD. Many empirical models \cite{Betz1972charge,shima1982empirical, SHIMA1986357, SHIMA1992173, schiwietz2001improved,schiwietz2004femtosecond} have been used to obtain such an equilibrated $q^o_m$. Of these models, the \citet{schiwietz2001improved} model which is even used in recent times \cite{PhysRevA.108.052817, chatterjee2021significance, chatterjee2021exploring} for the prediction of $q_m^{o}$ through solid target foils is named the SGM. Note that after a few years of the appearance of SGM, it was improved by the same group \cite{schiwietz2004femtosecond}, we call it improved SGM (ISGM). However, no description of CSD with this empirical model is found there. If we consider the distribution width formula given in \citet{schiwietz2001improved} remains valid here too, but that formula contains the charge state fractions (F(q)) as follows
\begin{equation}
\Gamma=[\sum_q (q-q_m^{o})^2 F(q)]^{1/2}
\label{d}
\end{equation}
Hence, these formulae can only be used to calculate the CSD if F(q) is also known and it is possible from the experimental data only. Specifically, the width parameter here makes it obvious that the model cannot predict F(q) independently. This dependency on experimental data restricts the predictive power of ISGM \cite{schiwietz2004femtosecond}, making it ineffective for applications where a complete charge-state distribution is required. Hence, neither SGM nor ISGM gives any description of CSD as they do not directly address the determination of the charge state distribution (CSD), which is essential for many practical applications. Although theoretical models such as ETACHA \cite{JPRozet_1989, ROZET199667, ETACHA4, SWAMI2018120}, CHARGE (a three-state model), and GLOBAL (a 28-state model) \cite{scheidenberger1998charge} have been developed to obtain CSD to simulate charge-exchange dynamics, they also have limitations. ETACHA \cite{JPRozet_1989, ROZET199667, ETACHA4, SWAMI2018120} works effectively for intermediate and high-energy regimes \cite{ROZET199667}, but struggle with heavy ions that have more electrons than Ni-like ions. CHARGE and GLOBAL \cite{scheidenberger1998charge}, on the other hand, are primarily designed for high-energy relativistic ions, leaving a gap in the prediction of charge-state distributions for the intermediate energies. These limitations require a reconsideration of earlier empirical models and exploration of a new approach to obtain CSD independently (without requiring any assistance from experimental charge-state fractions).

Heavy ions with intermediate energies are quite easy to produce and are very useful for accelerator applications, as heavy ions in the intermediate energy range are used to pass through the charge stripper to have higher charge states, so that higher energy will be obtained from the booster stage of the large accelerators. On this ground, empirical formulae may find their value if we find a way to generate CSD using the estimated values $q_m^{o}$. Normally, $q_m^{o}$ is the center point of the CSD, which means that the highest $F(q)$ is obtained at the rounded value of $q_m^{o}$. However, during practical applications, a full CSD curve is very essential if an accelerator has to use any other charge states to obtain an ion beam of the required energy for a certain application. Hence, a methodology for estimating the $F(q)$ without limitations is extremely important. In this respect, empirical models find their worth. 
\par
To estimate the CSD, many empirical models exist in the literature \cite{nikolaev1968equilibrium,betz1970charge, Betz1972charge}. These models used $q_m^{o}$ and $\Gamma$ in the Gaussian distribution to obtain the charge state fraction. Different formulae are used to obtain the correct $q_m^{o}$ for different regions of the atomic numbers of the projectile \cite{Betz1972charge}. To eliminate such difficulties, \citet{schiwietz2001improved} and \citet{schiwietz2004femtosecond} formulated a single formula to empirically calculate $q_m^{o}$ for the entire range of projectiles and targets. 

\begin{figure}
\includegraphics[width=8.2cm,height=8.2cm]{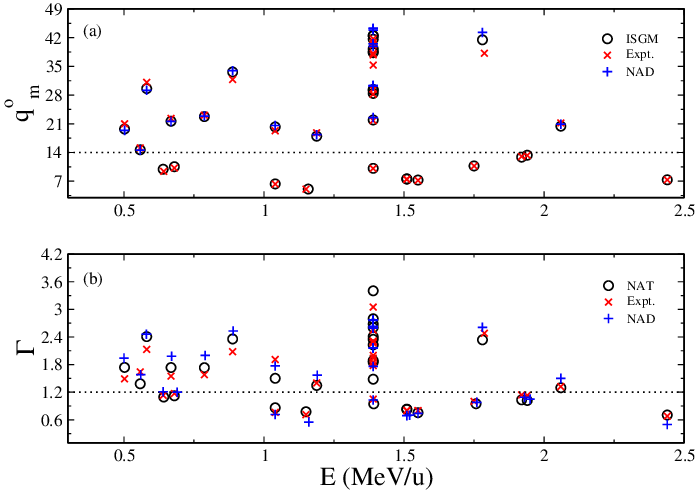}
\caption{(a) Projectile mean charge state ($q_m^{o}$) versus beam energy (E(MeV/u)): Comparison of experimental $q_m^{o}$ (red cross) \cite{SHIMA1986357, BALL1990125} with the empirical $q_m^{o}$ using ISGM (black circle) \cite{schiwietz2004femtosecond} and NAD (blue plus) \cite{nikolaev1968equilibrium}.
(b) Distribution width ($\Gamma$) versus beam energy (E(MeV/u)): Comparison of experimental $\Gamma$ (Red cross) \cite{SHIMA1986357, BALL1990125} with empirical $\Gamma$ using NAT (black circle) \cite{novikov2014methods} and NAD (blue plus) \cite{nikolaev1968equilibrium}.}
\label{qm and distribution}
\end{figure}

\par
In this work, we attempt to estimate (i) $q_m^o$ in good agreement with the experimental values, (ii) the distribution width close to the experimental figures and (iii) finally $F(q)$ and CSD in accordance with the experiments so that these estimated values can be used in different fundamental and practical applications. We present here the methodologies and other details.
\section{CHARGE STATE DISTRIBUTION OF THE PROJECTILE ION OUTSIDE THE TARGET FOIL}
Out of several models mentioned above, the Schiwietz Grande Model (SGM) \cite{schiwietz2001improved} was used mainly to predict the mean charge state ($q_m^{o}$) because a single formula worked for the entire range of projectile ions on solid target foils. However, it often does not represent the experimental scenario. Consequently, the same group developed an improved formula (ISGM) \cite{schiwietz2004femtosecond}. We used ISGM to calculate $q_m^{o}$. According to ISGM , the $q_m^{o}$ is expressed as a function of a reduced parameter $x_o$ and the projectile atomic number $Z_1$ as follows
\begin{equation}
q_m^{o} = \frac{Z_1 \left(8.29 \, x_o + x_o^4\right)}{\frac{0.06}{x_o} + 4 + 7.4 \, x_o + x_o^4}
\label{qmo'}
\end{equation}

\begin{figure}
\includegraphics[width=8cm,height=6cm]{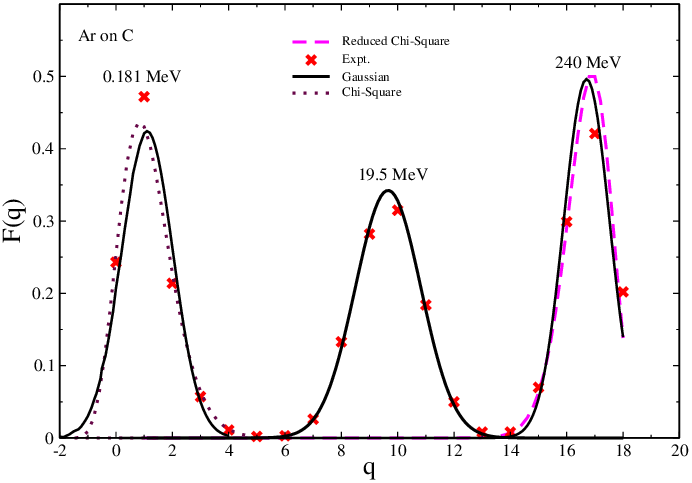}
\caption{Charge state distribution (CSD) curve of Ar ions on carbon target foil (Charge state fraction (F(q)) versus charge state (q)): Comparison of the experimental CSD (red cross) with $\chi^2$ distribution (dotted maroon) for low charge Ar ions, reduced-$\chi^2$ distribution (dash pink) for high-charge Ar ions and  Gaussian distribution (black) for entire charge range of Ar ions. }
\label{shima}
\end{figure}

\begin{figure*}
\includegraphics[width=16cm,height=12cm]{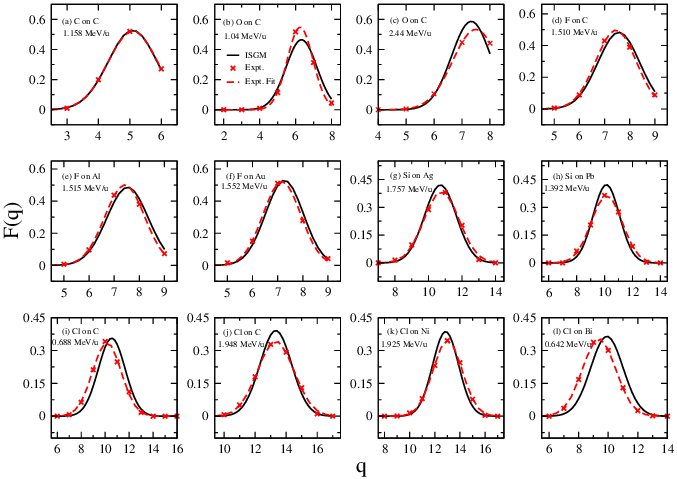}
\caption{Gaussian charge state distribution (CSD) curves using $q^o_m$ from ISGM \cite{schiwietz2004femtosecond} and distribution width from NAT \cite{novikov2014methods} for different projectile ions ($Z_1\le17$) and various solid targets at diverse energies are compared with the corresponding experimental results.}
\label{light}
\end{figure*}
\begin{figure*}
\includegraphics[width=17cm,height=15cm]{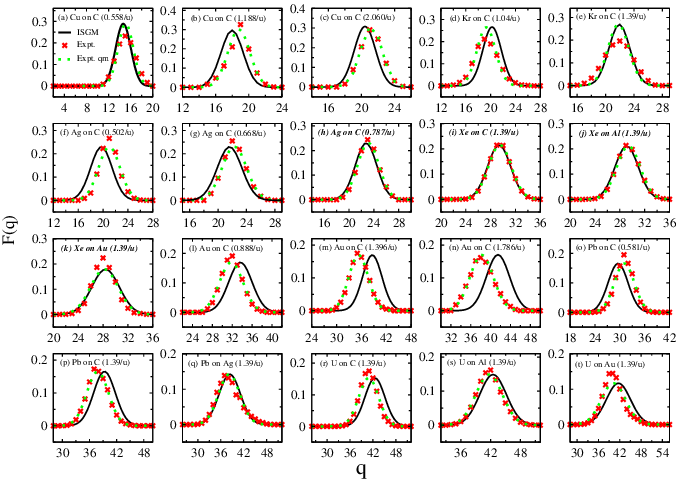}
\caption{Gaussian charge state distribution (CSD) curves with the distribution width from NAT \cite{novikov2014methods} for different projectile ions ($Z_1>17$) and various solid targets at diverse energies using $q_m^{o}$ from (i) ISGM (black) \cite{schiwietz2004femtosecond} and (ii) experiments (green) \cite{SHIMA1986357, BALL1990125} are compared with the corresponding experimental results.}
\label{heavy}
\end{figure*}

Here the reduced parameter $x_o$ is written in terms of projectile velocity ($v_1$), projectile atomic number $Z_1$, and target atomic number $Z_2$ as follows
\begin{equation}
x_o = c_1 \left(\frac{\bar{v}}{c_2 \cdot 1.54}\right)^{\left(1 + \frac{1.83}{Z_1}\right)}
\label{xo}
\end{equation}
\begin{equation}
\bar{v} = Z_1^{-0.543} \left(\frac{v_1}{v_o}\right)
\end{equation}
\begin{equation}
c_1 = 1 - 0.26 \exp\left(-\frac{Z_2}{11}\right) \exp\left(-\frac{(Z_2 - Z_1)^2}{9}\right)
\end{equation}
\begin{equation}
c_2 = 1 + 0.030 \, \bar{v} \, \ln(Z_2)
\end{equation}

where $v_o$ is the Bohr velocity $(2.19\times10^6 m/s)$.
Next, the distribution width $\Gamma$ is taken from  \citet{novikov2014methods} (NAT) as
\begin{equation}
\centering
\Gamma(x_1)= C[1-exp(-(x_1)^\alpha)][1-exp(-(1-x_1)^\beta)]
\label{Gamma}
\end{equation}
where $x_1$ is given as
\begin{equation}
x_1= q_m^{o}/Z_1   
\label{x_1}
\end{equation}
and \:$\alpha=0.23$,\: $\beta=0.32$, $C=2.669-0.0098.Z_2+0.058.Z_1+0.00048.Z_1.Z_2$.
\begin{figure}
\includegraphics[width=8.2cm,height=8.2cm]{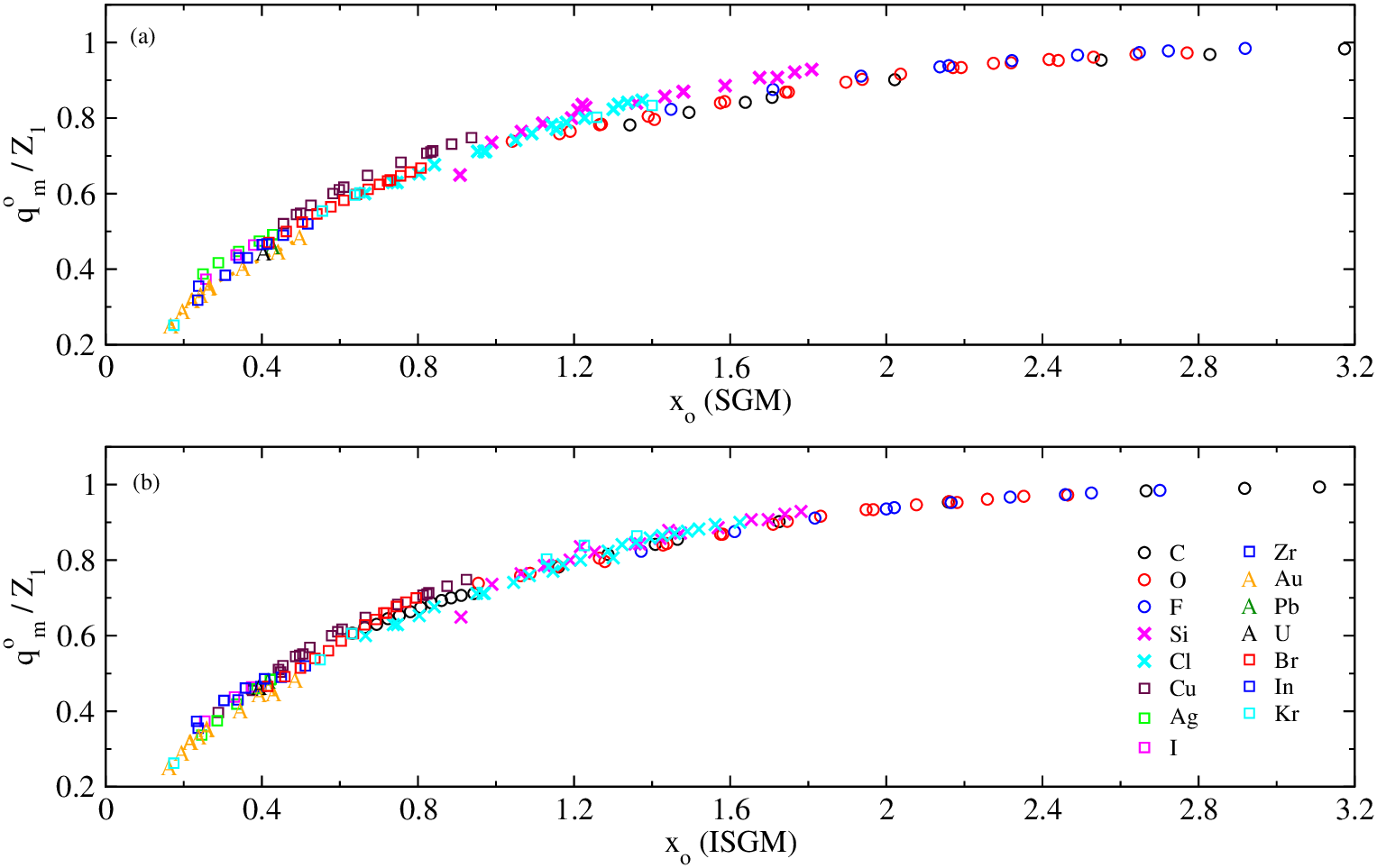}
\caption{Experimental $q_m/Z_1$ versus $x_o$ plot: the reduced parameter $x_o$ is taken from \citet{schiwietz2001improved} for (a) and \citet{schiwietz2004femtosecond} for (b). Experimental $q_m^o$ data for different projectiles on carbon targets are taken from Ref.\cite{SHIMA1986357, BALL1990125}.}
\label{Distribution}
\end{figure}

Next, we have compared the mean charge state $q_m^{o}$ as obtained from ISGM \cite{schiwietz2004femtosecond} and the width of the charge state distribution $\Gamma$ from \citet{novikov2014methods} (NAT) with the corresponding experimental data \cite{SHIMA1986357, BALL1990125} as a function of the energy of emergence in Fig.\ref{qm and distribution}. The exit energy is estimated by deducting the ion energy loss in the foil calculated using the SRIM package \cite{ziegler2010srim} from the incident beam energy.  Here, one can see that both these $q_m^{o}$ and $\Gamma$ agree well with the experimental data for the lighter heavy ions up to $Z_1=16$. However, the picture is not the same for the heavy ions $Z_1>16$. To check whether \citet{nikolaev1968equilibrium} (NAD) formula for predicting the mean charge state of projectile ions through solid targets ($q_m^{o}$) suits better, we have used the $q_m^{o}$ ($Z_1>16$) as 
\begin{equation}
q_m^{o}=Z_1[1+(Z{_1}^{-0.45} v_1/v')^{-1/0.6)}]^{-0.6}
\label{w}
\end{equation}
here $v'$= $3.6 \times 10^{8}$ cm/sec.

\begin{figure}
\includegraphics[width=8.2cm,height=8.2cm]{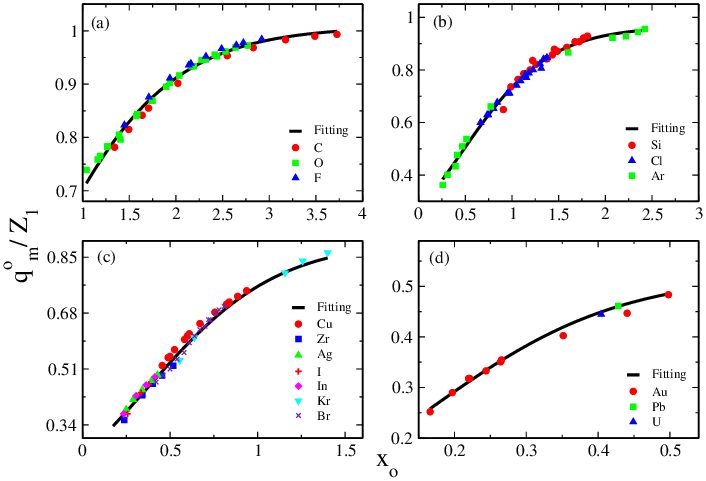}
\caption{Same as Fig. \ref{Distribution} but the data are divided into four groups: data for ion beams (a) $Z_1 \le 10$, (b)  $10 < Z_1 \le 18$, (c)  $18 < Z_1 \le 54$ and (d)  $54 < Z_1 \le 92$.}
\label{Fig2}
\end{figure}


\begin{figure}
\includegraphics[width=8.2cm,height=11.5cm]{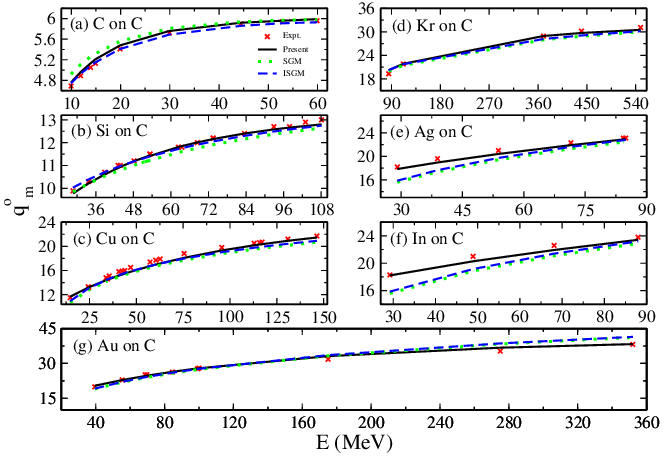}
\caption{Comparison of experimental mean charge states ($q_m^o$) of different ion beams on carbon target \cite{SHIMA1986357, BALL1990125} with SGM \cite{schiwietz2001improved}, ISGM \cite{schiwietz2004femtosecond}, and our model as a function of beam energy (E).}
\label{QmE}
\end{figure}

\begin{figure}
\includegraphics[width=8.2cm,height=8.2cm]{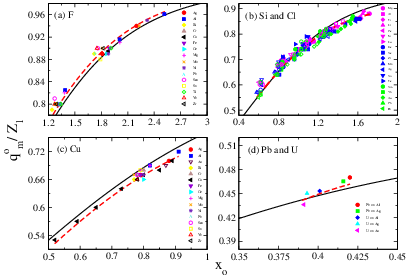}
\caption{Same as Fig. \ref{Fig2} but for miscellaneous targets excepting the carbon 
for different projectile ions. Solid lines represent Eqns 16-19 for Figs.(a) to (d), respectively and the dash lines is drawn with respect to the equations  given in Table 1. }
\label{Fig3}
\end{figure}

\begin{figure}
\centering
\includegraphics[width=8.2cm,height=11cm]{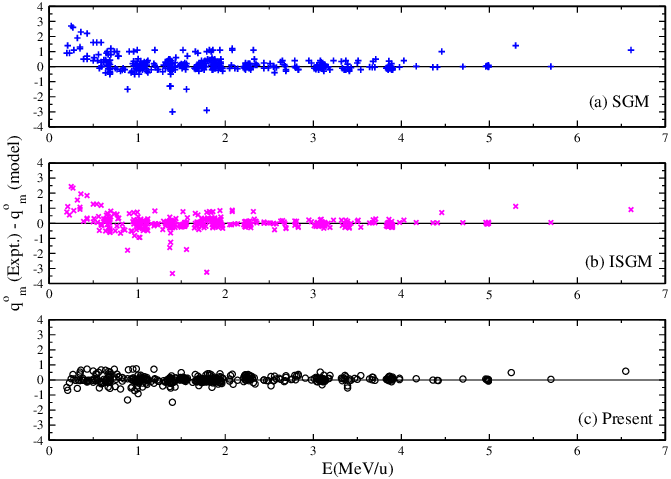}
\caption{Deviation of mean charge state predictions of different models from the corresponding experimental mean charge states for different ion beams at different energies on different target foils: (a) $q_m^o$ (Expt.) - $q_m^o$ (SGM), (b) $q_m^o$ (Expt.) - $q_m^o$ (ISGM) and (c) $q_m^o$ (Expt.) - $q_m^o$ (present).} 
\label{difference}
\end{figure}

\begin{figure}
\includegraphics[width=8.2cm,height=11.0cm]{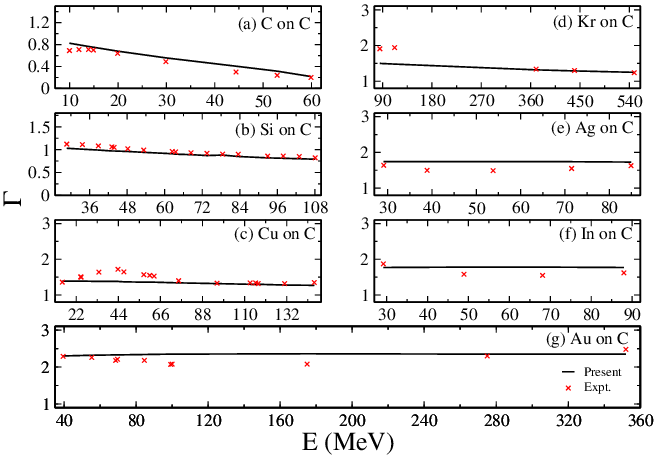}
\caption{Comparison of experimental charge state distribution width ($\Gamma$) \cite{SHIMA1986357} of $^{12}$C, $^{28}$Si, $^{63}$Cu, $^{83}$Kr, $^{107}$Ag, $^{114}$In, and $^{197}$Au ion beams on carbon targets with theoretical values calculated by Eqn.\ref{Gamma} \cite{novikov2014methods} as a function of beam energies. }
\label{GAMMA}
\end{figure}

\par
\begin{table}[ht]
\centering
\renewcommand{\arraystretch}{0.9}
\caption{Modification of mean charge state $q^o_{m_i}$ (i=1 to 4) for ion beams on heavy targets ($Z_2 \ge 12)$.}
\label{tab:my_table}
\begin{tabular}{|c|c|c|c|}
\hline
\textbf{\(Z_1\)} & \textbf{\(Z_2\)} & \textbf{\(E\) (MeV/u)} & \textbf{Final \(q^o_m\)} \\ \hline
\(Z_1 \le 10\) & \(12 \le Z_2 \le 39\) & \(0<E\le4\) & \(q^o_{m_1}\) \\ \hline
\(Z_1 \le 10\) & \(40 \le Z_2 \le 54\) & \(E > 1.4\) & \(q^o_{m_1} + {0.8\,Z_1}/{Z_2}\) \\ \hline
\(Z_1 \le 10\) & \(54 < Z_2 \le 92\) & \(E > 1.4\) & \(q^o_{m_1} + {1.5\,Z_1}/{Z_2}\) \\ \hline

\(11\le Z_1\le 18\) & \(12 \le Z_2 \le 18\) & \(0<E\le1\) & \(q^o_{m_2} + {0.2\,Z_1}/{Z_2}\) \\ \hline
\(11\le Z_1\le 18\) & \(19 \le Z_2 \le 69\) & \(0<E\le1\) & \(q^o_{m_2} + {0.7\,Z_1}/{Z_2}\) \\ \hline
\(11\le Z_1\le 18\) & \(70 \le Z_2 \le 92\) & \(0<E\le1\) & \(q^o_{m_2} - {1.4\,Z_1}/{Z_2}\) \\ \hline

\(11\le Z_1\le 18\) & \(12 \le Z_2 \le 18\) & \(1<E\le1.5\) & \(q^o_{m_2} + {0.05\,Z_1}/{Z_2}\) \\ \hline
\(11\le Z_1\le 18\) & \(19 \le Z_2 \le 46\) & \(1<E\le1.5\) & \(q^o_{m_2} + {0.2\,Z_1}/{Z_2}\) \\ \hline
\(11\le Z_1\le 18\) & \(47 \le Z_2 \le 69\) & \(1<E\le1.5\) & \(q^o_{m_2} + {1.2\,Z_1}/{Z_2}\) \\ \hline
\(11\le Z_1\le 18\) & \(70 \le Z_2 \le 92\) & \(1<E\le1.5\) & \(q^o_{m_2} - {0.1\,Z_1}/{Z_2}\) \\ \hline

\(11\le Z_1\le 18\) & \(12 \le Z_2 \le 18\) & \(1.5<E<3.11\) & \(q^o_{m_2} - {0.05\,Z_1}/{Z_2}\) \\ \hline
\(11\le Z_1\le 18\) & \(19 \le Z_2 \le 47\) & \(1.5<E\le3.11\) & \(q^o_{m_2} - {0.2\,Z_1}/{Z_2}\) \\ \hline
\(11\le Z_1\le 18\) & \(48 \le Z_2 \le 69\) & \(1.5<E\le3.11\) & \(q^o_{m_2} + {1.2\,Z_1}/{Z_2}\) \\ \hline
\(11\le Z_1\le 18\) & \(70 \le Z_2 \le 92\) & \(1.5<E \le 3.11\) & \(q^o_{m_2} - {1.2\,Z_1}/{Z_2}\) \\ \hline

\(11\le Z_1\le 18\) & \(12 \le Z_2 \le 31\) & \(3.11\le E\le4\) & \(q^o_{m_2} - {0.3\,Z_1}/{Z_2}\) \\ \hline
\(11\le Z_1\le 18\) & \(32 \le Z_2 \le 70\) & \(3.11<E\le4\) & \(q^o_{m_2} - {Z_1}/{Z_2}\) \\ \hline
\(11\le Z_1\le 18\) & \(70 < Z_2 \le 92\) & \(3.11<E\le4\) & \(q^o_{m_2} - {2\,Z_1}/{Z_2}\) \\ \hline

\(19\le Z_1\le 54\) & \(12 \le Z_2 \le 27\) & \(0\le E\le4\) & \(q^o_{m_3} - {0.1\,Z_1}/{Z_2}\) \\ \hline
\(19\le Z_1\le 54\)\ & \(28 \le Z_2 \le 69\) & \(0\le E\le4\) & \(q^o_{m_3} - {0.4\,Z_1}/{Z_2}\) \\ \hline
\(19\le Z_1\le 54\)\ & \(70 \le Z_2 \le 92\) & \(0\le E\le4\) & \(q^o_{m_3} - {Z_1}/{Z_2}\) \\ \hline

\(54\le Z_1\le 92\)\ & \(12 \le Z_2 \le 47\) & \(0\le E\le4\) & \(q^o_{m_4} + 1.5\,{Z_2}/{Z_1}\) \\ \hline
\(54\le Z_1\le 92\)\ & \(48 \le Z_2 \le 69\) & \(0\le E\le4\) & \(q^o_{m_4} + {1.2\,Z_2}/{Z_1}\) \\ \hline
\(54\le Z_1\le 92\)\ & \(70 \le Z_2\le 92\) & \(0\le E\le4\) & \(q^o_{m_4} - 0.5 \,{Z_2}/{Z_1}\) \\ \hline
\end{tabular}
\label{Assumptions}
\end{table}

\begin{figure}
\includegraphics[width=8.3cm,height=11cm]{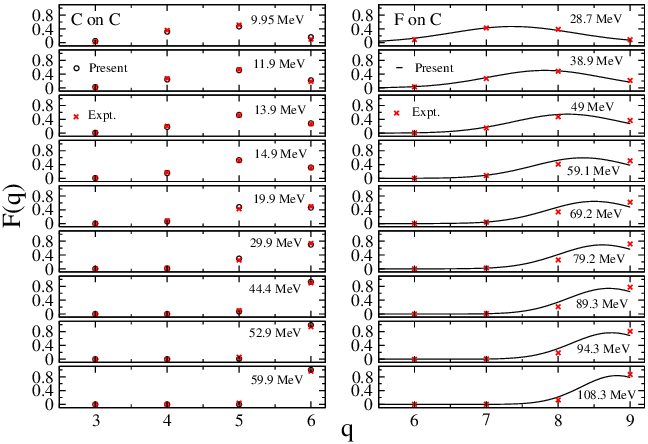}
\caption{Comparison of experimental charge state fraction F(q) vs q for $^{12}$C and $^{19}$F ion beams at various beam energies \cite{SHIMA1986357} with our theoretical values where F(q) for a particular q is calculated using the Gaussian distribution using Eqn.\ref{FQ} with distribution width from Eqn. \ref{Gamma} and $q_m^o$ from Eqn. (\ref{a}).}
\label{light1}
\end{figure}

\begin{figure}
\includegraphics[width=8.3cm,height=11cm]{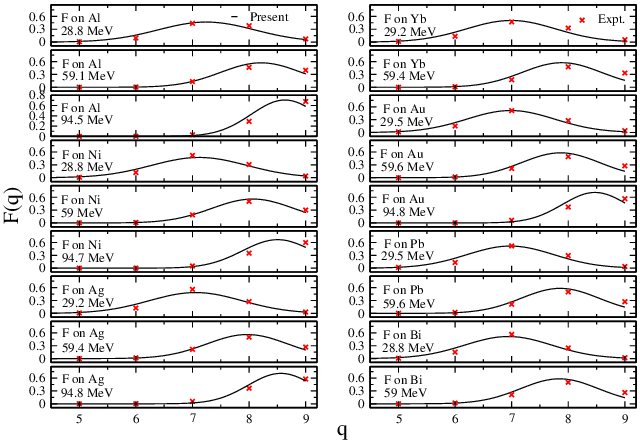}
\caption{Same as Fig.\ref{light1} but for $^{19}$F beam on different target foils.}
\label{light2}
\end{figure}

\begin{figure}
\includegraphics[width=8.3cm,height=12cm]{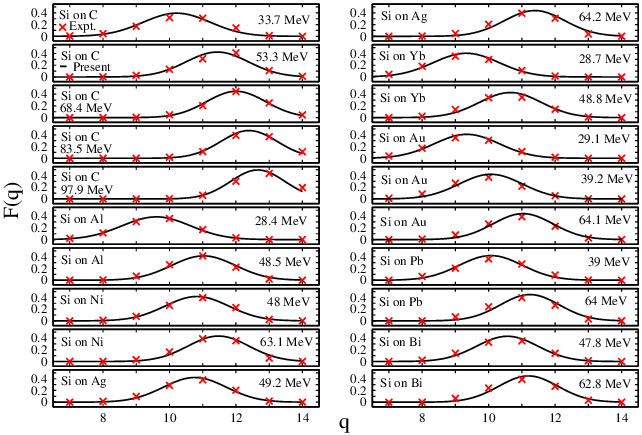}
\caption{Same as Fig.\ref{light1} but for $^{28}$Si beam on different target foils.}
\label{Si-C-All}
\end{figure}

\begin{figure}
\centering
\includegraphics[width=8.3cm,height=12cm]{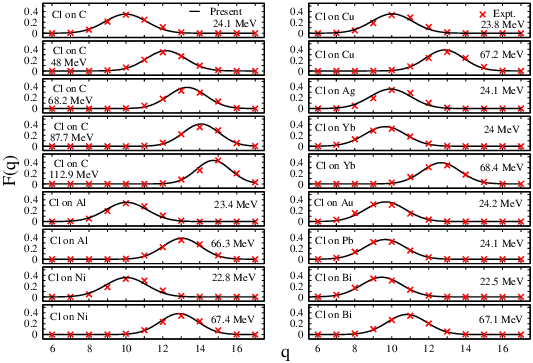}
\caption{Same as Fig.\ref{light1} but for $^{35}$Cl beam on different target foils.}
\label{Cl-C-All}
\end{figure}

\begin{figure}
\centering
\includegraphics[width=8.3cm,height=12cm]{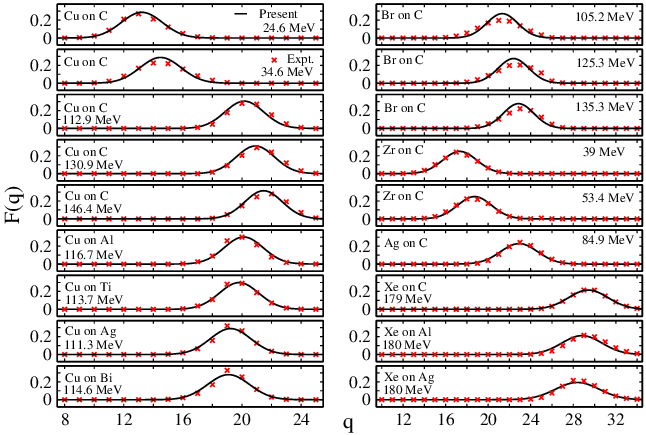}
\caption{Same as Fig.\ref{light1} but for $^{63}$Cu,  $^{80}$Br, $^{91}$Zr, $^{107}$Ag and $^{131}$Xe beam on different target foils.}
\label{Cu to Xe}
\end{figure}

\begin{figure}
\centering
\includegraphics[width=8.3cm,height=11cm]{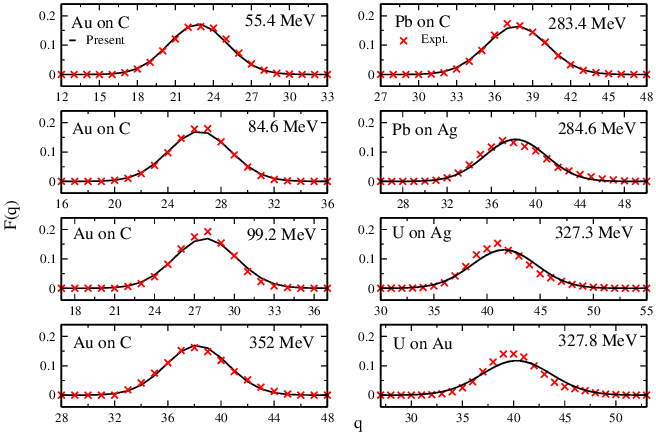}
\caption{Same as Fig.\ref{light1} but for $^{197}$Au, $^{207}$Pb and $^{238}$U beam on different target foils.}
\label{Pb and U}
\end{figure}

\par
In addition to the above formula, many other formulas that work for different $Z_1$ are also given in the review of \citet{Betz1972charge}. The agreement of $q_m^{o}$ in ISGM (black) \cite{schiwietz2004femtosecond} and NAD (blue) \cite{nikolaev1968equilibrium} with the experimental mean charge state \cite{SHIMA1986357, BALL1990125} of projectile ions through solid targets (red cross) seems to look quite similar. To examine which one of these agrees better in this comparison, we have tested the data by using a quantity called relative sum metrics (RSM). RSM is just the residual sum square (RSS=$\sqrt{\sum_i^n [y_i - f(E_i)]^2}$) divided by the sum of squares of the experimental values ($y_{i}$) as follows:

\begin{equation}
RSM =  \frac{\sqrt{\sum_i^n [y_i - f(E_i)]^2}}{\sum_i^n {y_i}},
\end{equation}

\par
where f($E_i$) is the theoretical counterpart of $y_i$. Comparing the RSM obtained from two models, we saw that ISGM \cite{schiwietz2004femtosecond} gives an error/RSM of 0.010 while NAD \cite{nikolaev1968equilibrium} shows an error of 0.018 for $q_m^{o}$ (Fig.1). Similarly, the distribution width $\Gamma$ for different projectile $Z_1$ is also given by \citet{nikolaev1968equilibrium} as follows
\begin{equation}
\Gamma=0.5[q_m^{o}(1-(q_m^{o}/Z_1)^{1/0.6})]^{1/2}
\label{fx}
\end{equation}
Using $q_m^{o}$ from Eq.\ref{qmo'} we further test the case of two distribution widths of NAT \cite{novikov2014methods} and NAD \cite{nikolaev1968equilibrium}. We found that Eq.\ref{Gamma} or $\Gamma$ of \citet{novikov2014methods} gives an error of 0.023 and Eq.\ref{fx} or NAD has an error of 0.028. Hence, we have used ISGM \cite{schiwietz2004femtosecond} for $q^o_m$ and NAT \cite{novikov2014methods} for $\Gamma$.

Next, to get the charge state fractions (F(q)), we have used the Gaussian charge state distribution \cite{SHIMA1986357} as
\begin{equation}
F(q) = \frac{1}{ \sqrt{2\pi}}\frac{1}{\Gamma} exp(-(q-q_m^{o})^2 / 2 \Gamma^2) 
\label{FQ}
\end{equation}
provided
\begin{equation}
\sum_q F(q)=1. 
\end{equation}

\noindent However, \citet{SHIMA1986357} used three different distributions such as the $\chi^2$, Gaussian, and reduced $\chi^2$ distributions for CSD at low, intermediate, and high charge projectile ions, respectively. In Fig.\ref{shima}, we have compared the Gaussian charge state distribution for the entire range of charge states (low, intermediate, and high) with the three distributions for Ar-ions on the carbon target, as shown in \citet{SHIMA1986357}. We examined the goodness-of-fit for all distributions using a measure of RSM studies. The RSM with the Gaussian distribution turns out to be 0.0803, 0.0204, and 0.000087 for low-, intermediate-, and high-charge Ar ions, respectively. The RSM is 0.0466 and 0.000101 for the $\chi^2$ (low charge) and reduced-$\chi^2$ (high charge) distributions, respectively. It implies that the $\chi^2$ distribution for low charge is certainly better than the Gaussian charge state distribution curve with $q^o_m$ from ISGM \cite{schiwietz2004femtosecond} and $\Gamma$ from NAT \cite{novikov2014methods} in fitting the experimental data \cite{SHIMA1986357, BALL1990125}. However, the latter is good not only for intermediate-charge Ar ions but also for high-charge Ar ions. Hence, the use of $\Gamma$ from \citet{novikov2014methods} in the Gaussian charge state distribution curve (Eqn.\ref{FQ}) may address both intermediate and high-charge projectile ions well.

\section{PROBLEM ARISES WITH ISGM $q_m^{o}$}

\par
To test the above-mentioned methodology, different projectile ($Z_1\le17$) target systems at various energies have been shown in Fig.\ref{light}. We see that the empirical formula for F(q) fits well with the experimental data \cite{SHIMA1986357, BALL1990125} except for two cases (i) and (l). In both cases, Cl is the projectile, while carbon and bismuth are the targets in (i) and (l), respectively. Note that Cl is the heaviest projectile of all the projectiles used in this figure. Four cases with the Cl projectile (i-l), (j), and (k) are also shown with different targets of carbon and nickel, respectively, but with higher beam energies. In contrast, (i) and (l) are done with lower beam energies. Furthermore, in Fig.\ref{qm and distribution}, we have already mentioned that $q_m^{o}$ is not good for heavy ions.  Hence, the reason for the disagreement in CSD is both the low beam energy and heavier projectiles. 
 
\par
To clarify the above-mentioned doubts, we have investigated the heavy-ion cases in Fig.\ref{heavy}. Except for a few cases in Fig.\ref{heavy} - (h), (i), (k), the other 17 data sets show a clear mismatch between the empirical and experimental CSD. To find the real cause of this disagreement, we employed the same method using $\Gamma$ from \citet{novikov2014methods} and the Gaussian charge state distribution from Eqn.\ref{FQ}, but the mean charge state of the projectile ion is taken from experimental data \cite{SHIMA1986357, BALL1990125}. Using this technique, we found good agreement between the empirical (green curve) and experimental (red cross) charge-state distribution even with low beam energies. Hence, there is no issue with the proposed empirical formulation as far as the distribution width and Gaussian distribution are concerned in representing the experimental charge state distributions. Our challenge remains to evaluate the mean charge state outside the foil well for the heavier ions ($Z_1\ge17$). The next attempt was to find the mean charge state also empirically instead of experimental ones. A thorough search shows that the mean charge state obtained from \citet{schiwietz2004femtosecond} agrees well with the experimental results if $Z_1 \le16$, but the scenario is not so for $Z_1 > 16$ as we see from the above results. So far, this work concludes that the challenge of an empirical formula for evaluating the mean charge states outside the target foil is not yet over. Hence, to improve $q_m^{o}$ empirically, we proceed far to solve this problem.

\section{Renewed formulae for $q_m^o$}
Improvement of ISGM from SGM was done through the reduced parameter $x_o$. To investigate the role of $x_o$ in the mean charge state formula, we examined the relationship between the parameter $x_o$ and the normalized charge state $q_m^o$/$Z_1$ for both the SGM and ISGM models in Fig.\ref{Distribution}. The parameter $x_o$ for SGM \cite{schiwietz2001improved} is given as
\begin{equation}
x_o =(v_1/v_o Z_1^{-0.52} Z_2^{-0.019Z_1^{-0.52} {v_1/v_o}}/1.68)^{1+1.8/Z_1}
\label{xoSGM}
\end{equation}
We revealed that Fig.\ref{Distribution}(a) exhibits a distinctive four-band pattern in the data, which is not observed in Fig.\ref{Distribution}(b).  Exactly, this was the objective of ISGM. However, this upgrade did not work well, so we decided to use the earlier $x_o$, in different perspectives, as defined in the SGM (Eq. \ref{xoSGM}) from this instance.
\par
The reason for this consideration of the earlier $x_o$ parameter is quite clear from Fig. \ref{Distribution}(a); the data constitute a wide-spread band, and a single fitted line is too simple to provide fairness to such data. Furthermore, we notice that projectiles on heavier targets ($Z_2 > 6$) particularly widen the width of the curve. To avoid such an issue, we first consider only the experimental data on the light (carbon) targets, which can be classified into four groups of projectile ions as follows: (a) $Z_1 \le 10$, (b) $10 < Z_1 \le 18$, (c) $18 < Z_1 \le 54$, and (d) $54 < Z_1 \le 92$. The $q_m/Z_1$ are plotted as a function of the same reduced parameter $x_o$ as given in Eqn. (\ref{xoSGM}) and all the four curves have been fitted with the standard logistic function as follows
\begin{equation}
\frac{q_m^{o}}{Z_1} = \frac{L}{1+exp(-K(x_o-x'))}
\label{general}
\end{equation}
where $x'$ is the midpoint of the $x_o$ value, L is the supremum of the values of the function and K is the logistic growth rate of the curve. The $q_m^o$ for the projectile ions $Z_1 \le 10$ (call it $q^o_{m_1}$) can be obtained by using the empirical formula as follows
\begin{equation}
q^o_{m_1} = Z_1 \frac{1.007704}{1+exp(-1.419686(x_o-0.416069))}.
\label{a}
\end{equation}
\par
Similarly, the $q_m^o$ for the projectile ions  $10 < Z_1 \le 18$ (call it $q^o_{m_2}$), $18 < Z_1 \le 54$ (call it $q^o_{m_3}$) and $54 < Z_1 \le 92$ (call it $q^o_{m_4}$) can be obtained from the following empirical formulae, respectively.
\begin{equation}
q^o_{m_2} = Z_1 \frac{0.966005}{1+exp(-2.117365(x_o-0.454663))},
\label{b}
\end{equation}
\begin{equation}
q^o_{m_3} = Z_1 \frac{0.901506} {1+exp(-2.674817(x_o-0.368080))}
\label{c}
\end{equation}
and
\begin{equation}
q^o_{m_4} = Z_1 \frac{0.525255}{1+exp(-7.609263(x_o-0.170544))}.
\label{d'}
\end{equation}
\par
There is quite a sufficient number of data for fitting the curve of Fig. \ref{Fig2}(a)-(c), but the data available for fitting the curve of Fig. \ref{Fig2}(d) are not many. Only data exist with Au at several energies and a few points for Pb and U projectiles. In spite of this, we got a good fit, and the corresponding equation can be used for our future use. Furthermore, two data points for gold ions deviate from the fitted line; the shell effect may be one of the reasons. 
\par
In the above equations, $q_m^o$ varies from $q^o_{m_1}$ to $q^o_{m_4}$ depending on the projectile ions. Next, to obtain the values of the charge state fraction (F (q)), we have used the Gaussian charge state distribution from Eq.\ref{FQ} and $\Gamma$ is taken from the distribution width given by \citet{novikov2014methods} (Eq. \ref{Gamma}). We replace the quantity $x_1$ as used in Eq.\ref{x_1} to calculate the width distribution $\Gamma (x_1)$ in Eq. \ref{Gamma} as follows
\begin{equation}
x_1 = \frac{q^o_m(m=m_1, m_2, m_3, m_4)}{Z_1}
\label{x_1new}
\end{equation}

\section{Results and discussions} 
We have used the reduced parameter $x_o$ of the SGM from equation \ref{xoSGM} and the ratio of $q_m^o$ and $Z_1$ plotted along the x-axis and y-axis, respectively, in Fig.\ref{Distribution}(a). Then it was observed that all the data do not show the same trend. A clear difference can be seen here as we move from $Z_1$ = 6 to 92. Hence, we divide the entire existing data into four categories according to their trend and fit it accordingly in Fig.\ref{Fig2}. This fitting gives us a clear picture to obtain $q_m^o$ well and yields the fitting equations \ref{a} to \ref{d'} for any ion beams on the carbon target. A representative comparison of experimental $q_m^o$ (red cross) for different ion beams on carbon target foil with SGM (dotted green), ISGM (dash blue), and present model (solid black line) is presented in Fig.\ref{QmE}. It shows clearly that the present model best agrees with the experimental data. Furthermore, we know that ISGM is good for light ion regime $Z_1 \le 16$. To check whether present model predictions for $q_m^{o}$ in this region are even better than either SGM or ISGM, we have found out the RSM values of SGM, ISGM, and present model for two representative cases  (a)  and (b), i.e., for C and Si. RSM is found as 0.0046, 0.0023, and 0.0017 for SGM, ISGM, and the present formula for calculating $q_m^{o}$. RSM clearly indicated the usefulness of our proposed formulation for calculating $q_m^{o}$ even in the case of light-ion beams. However, the scenario with other than carbon targets as shown in Fig.\ref{Fig3} with solid line is not that good. To make it better, we have modified the equations \ref{a} to \ref{d'} by adding certain correction terms as given in Table \ref{Assumptions}. The comparison of the experimental data with such modified equations is shown with the dotted lines. A good agreement is observed between the experimental data and present estimates for all the groups.
\par
To make a comprehensive comparison between all the models, we have plotted the deviation of each model prediction from the experimental mean charge states as shown in Fig.\ref{difference}. We can clearly see from Fig.\ref{difference} that the scattering of data is the lowest. Therefore, we proceed to calculate the width distribution ($\Gamma$) using the present $q_m^{o}$ values rather than SGM or ISGM for all ion-solid collision processes.
\par
In the second step, the distribution width ($\Gamma$) of the projectile ion beam is calculated with an equation \ref{Gamma} taken from  \citet{novikov2014methods}, which corresponds to energy loss that occurred during the ion-solid collision process. Fig.\ref{GAMMA} represents the $\Gamma$ values of {$^{12}$C, $^{28}$Si, $^{63}$Cu, $^{83}$Kr, $^{107}$Ag, $^{114}$In and $^{107}$Au projectile ion beams for both theory and experimental values. Again we can see a fair agreement between the theory and the experimental data which suggests that $\Gamma$ is a good choice for this theory. It may be noteworthy here that in some of the above figures viz., Figs. \ref{qm and distribution}, \ref{QmE}, \ref{difference} and \ref{GAMMA}, either $q_m^o$ or $\Gamma$ were plotted as a function of emerging beam energies, which is estimated from the SRIM \cite{ziegler2010srim} code if not mentioned otherwise in the concerned reference. 

\par
In the third step, we have considered a Gaussian distribution from \citet{SHIMA1986357} and replaced its $\Gamma$ with $\Gamma$($x_1$) \cite{novikov2014methods} and use $q_m^o$ from Eqns.\ref{a}-\ref{d'} to obtain charge-state fraction values. An excellent agreement between theoretical and experimental CSD can be seen in Fig.\ref{light1}, which represents the graphs of F (q) versus q for the $^{12}$C and $^{19}F$ projectile ion beams on the carbon target at various beam energies. The same scenarios can be observed in Fig.\ref{light2} for the $^{19}F$ projectile ion beam on light to heavy targets foils, i.e. Al, Ni, Ag, Yb, Au, Pb, and Bi at different energies. It is very clear from Fig.\ref{light1} and Fig.\ref{light2} that the value of F (q) is the highest at a certain value of q. This implies that $q_m^o$ is quite close to this particular q. As the projectile ion beam increases, the mean charge state of the projectile ion shifts toward higher q values, which are much different from the incident charge state of the projectile ion beam.
\par
A similar observation can be seen for the ions in group II. For example, the cases of projectile ion beams $^{28}$ Si and $^{35}$ Cl are shown for different target foils at different energies in Fig.\ref{Si-C-All} and in Fig.\ref{Cl-C-All}, respectively. Here in both Fig.\ref{Si-C-All} and Fig.\ref{Cl-C-All}, excellent agreement is observed between the present model and the experimental data. This implies that both $q_m^{o}$ and $\Gamma(x_1)$ are predicted well by the present model.
\par
Now comes the heavier projectile ion beams lying in group III. We have shown the CSD of $^{63}Cu$, $^{80}$ Br, $^{91}$ Zr, $^{107}$ Ag, and $^{131}Xe$ on different target foils at different energies in Fig.\ref{Cu to Xe}. Here also, our predictions for CSD give very good agreement as shown for cases of $^{19}F$, $^{28}Si$, and $^{35}Cl$. 
\par
Now we proceed further towards much heavier projectile-ion beams lying in group IV. We see that the CSD for the heavy ions of $^{207}Pb$ and $^{238}U$ on the Ag and Au target foils, as shown in Fig. \ref{Pb and U}  gives an excellent match to the present predictions. 
\par 
From the above comparisons, it is confirmed that the choice of the correction factors introduced for the variation of the values $q_m^o$ with a heavier target is pretty good. However, incorporating so many corrections is quite difficult for practical applications. To circumvent such inconveniences we have planned to develop a code that will be reported appropriate journal shortly. 

\section{Conclusion}
The charge state distributions (CSD) of projectile ions through gaseous or solid targets in the energy range of tandem accelerators (1 MeV/u $<$ E $<$ 4 MeV/u) have a major impact on ion-atom collisions and accelerator physics. Theoretically, it is possible to calculate the CSD for up to Ni-like ions, but empirical models do not have such bounds. In recent decades, the mean charge states ($q_m^o$) have been obtained primarily from an empirical formula \cite{schiwietz2004femtosecond}. But no description of CSD is found there. To estimate the CSD, we have used a Gaussian distribution function having distribution width given by \citet{novikov2014methods}. The results obtained are compared with those obtained with the measured CSD for heavy projectile ions. We notice that it works well, but its validity range is restricted by the correct values of $q_m^o$, which are limited only to the projectile atomic number $Z_1 \le$ 16. Empirically, $q_m^o$ is estimated in terms of a reduced parameter $x_o$. Rather than using $x_o$  from \cite{schiwietz2004femtosecond}, we take it from \citet{schiwietz2001improved} and develop a new model, which classifies the full range of projectiles into four groups to estimate $q_m^o$ in good agreement with the experimental counterparts with the carbon targets. However, using this model to find $q_m^o$ for heavier targets shows a certain deviation from the experimental counterparts. To minimize this difference, we added certain correction terms with the formulae for $q_m^o$. The $q_m^o$ so obtained shows good agreement with the experimental scenarios. Using such $q_m^o$, we proceed to calculate the CSD in the above mentioned way. The CSDs thus obtained are found to be very close to the experimental values.

\section{Acknowledgement}
The authors thank Sanjiv Kumar for useful inputs.

\bibliography{1BIB.bib}
\bibliographystyle{apsrev4-1}
\end{document}